\documentclass[%
 reprint,
 amsmath,amssymb,
 aps,
 prb,
 floatfix
]{revtex4-2}

\usepackage{graphicx}
\usepackage{dcolumn}
\usepackage{bm}

\usepackage{comment}

\newcommand{\abs}[1]{\left|#1\right|}
\newcommand{\norm}[1]{\left\lVert#1\right\rVert}

\newcommand{\av}[1]{\left\langle#1\right\rangle}

\newcommand{\up}{\uparrow}
\newcommand{\dn}{\downarrow}

\renewcommand{\Re}{\operatorname{Re}}
\renewcommand{\Im}{\operatorname{Im}}

\usepackage{tikz}
\usetikzlibrary{calc,decorations,decorations.pathreplacing,shapes.geometric}

\definecolor{matrix5}{RGB}{255,255,178}
\definecolor{matrix4}{RGB}{254,217,118}
\definecolor{matrix3}{RGB}{254,178,76}
\definecolor{matrix2}{RGB}{253,141,60}
\definecolor{matrix1}{RGB}{240,59,32}
\definecolor{matrix0}{RGB}{189,0,38}

\begin{document}

\title{Two-particle correlations and the metal-insulator transition: Iterated Perturbation Theory revisited}

\author{Erik G. C. P. van Loon}
\affiliation{%
 Mathematical Physics Division,
 Department of Physics, Lund University, Lund, Sweden
}%

\begin{abstract}
Recent advances in many-body physics have made it possible to study correlated electron systems at the two-particle level. In Dynamical Mean-Field theory, it has been shown that the metal-insulator phase diagram is closely related to the eigenstructure of the susceptibility. So far, this situation has been studied using accurate but numerically expensive solvers. Here, the Iterated Perturbation Theory (IPT) approximation is used instead. Its simplicity makes it possible to obtain analytical results for the two-particle vertex and the DMFT Jacobian. The limited computational cost also enables a detailed comparison of analytical expressions for the response functions to results obtained using finite differences. At the same time, the approximate nature of IPT precludes an interpretation of the metal-insulator transition in terms of a Landau free energy functional.
\end{abstract}

\maketitle

\section{Introduction}

Electronic correlations play an important role in many phase transitions, including unconventional superconductivity, charge-density waves and magnetism. The seminal example of a correlation induced transition is the Mott metal-insulator~\cite{Imada98} transition. Once the repulsive Coulomb interaction between electrons overpowers the band energy, the electrons localize and the density of states at the Fermi level disappears, turning the system into an insulator. The minimal theoretical description of this transition is given by the dynamical mean-field theory solution of the Hubbard model in the limit of infinite dimension~\cite{Georges96}.

In addition to the density of states, the metal-insulator transition also has profound effects on level of two-particle correlation functions~\cite{Schafer13,vanLoon14,Schafer16}. In equilibrium statistical physics, these carry three simultaneous meanings. First, as the probabilty for creating and annihilating fermions at four specific times. Secondly, as the second derivative of the free energy with respect to a particular field, indicating the energy cost or gain associated with fluctuations. Thirdly, as the linear response of a one-particle observable to an external field. In dynamical mean-field theory, these three manifestations of two-particle correlations can be expressed entirely in terms of the local vertex~\cite{Khurana90,Kotliar00,Blumerphd,vanLoon20}. 

Over the last few years, two-particle correlations close to the metal-insulator transition have been a topic of active investigations~\cite{Krien19b,Melnick20,vanLoon20,Reitner20}, partially driven by methodological and computational improvements that have enabled the accurate determination of the two-particle vertex~\cite{Rohringer12,Hafermann12,Rigo22}. 
Despite these improvements, the study of two-particle correlations typically requires a substantial amount of supercomputing power, because the evaluation of the two-particle vertex across a wide range of frequencies is hard.

This manuscript takes a step back and considers the Iterated Perturbation Theory~\cite{Georges92b} approximation to dynamical mean-field theory. The vertex within this approximation is evaluated analytically and this makes it possible to study two-particle correlations in detail and with a very limited computational cost. In particular, IPT is very suited to the study of response functions as finite differences, since the method is free from stochastic noise.

The structure of this paper is as follows: Section~\ref{sec:overview} provides an overview of dynamical mean-field theory on the one- and two-particle level. This section summarizes the literature and translates results to the Bethe lattice where needed. In Section~\ref{sec:ipt}, IPT's two-particle vertex is derived and the resulting two-particle properties of DMFT+IPT are discussed. It is shown that the IPT approximation breaks certain symmetry properties of the many-body theory for the Hubbard model. This makes the construction of an IPT free energy functional problematic. Still, other two-particle quantities remain sensible. Section~\ref{sec:response} discusses the physical response to changes in the interaction strength. Using the results of the previous section, it is shown that $\Sigma(\tau)$ is a monotonous function of $U$ in IPT. Section~\ref{sec:numerics} then provides a numerical illustration of the derived analytical results. The paper ends with conclusions and an outlook.

\section{Dynamical mean-field theory on the Bethe lattice}
\label{sec:overview}

To set the stage, an overview of Dynamical Mean-Field Theory (DMFT) for the Hubbard on the Bethe lattice is given. More details can be found in the literature~\cite{Georges96,julichbook}. For the discussion of two-particle properties, the notation of Ref.~\cite{vanLoon20} will be followed, although it should be noted that some differences appear in the formulas: for the Bethe lattice considered here, the concept of momentum is absent. The lattice structure is encoded in the Bethe lattice self-consistency condition without invoking a momentum sum. 

\subsection{Hubbard Model}

The Hamiltonian of the Hubbard model~\cite{Hubbard63,Kanamori63,Gutzwiller63} is
\begin{align}
    H = -t \sum_{\av{ij}} \sum_{\sigma=\up,\dn} c^\dagger_{i\sigma} c^{\phantom{\dagger}}_{j\sigma} + U \sum_i n_{i\up} n_{i\dn}.
\end{align}
Here, $c^\dagger_{i\sigma}$ creates an electron with spin $\sigma$ at site $i$, $c_{i\sigma}$ is the corresponding annihilation operator and $n=c^\dagger c$ is a number operator, $\av{ij}$ denotes that the sum is taken only over pairs of neighboring sites, $t$ is the electron hopping and $U$ the electron-electron interaction. In this work, the Bethe lattice at half-filling is considered in the usual DMFT limit~\cite{Georges96} of $z\rightarrow \infty$ with $t \propto 1/\sqrt{z}$, where $z$ is the number of neighbors in the Bethe lattice. All calculations shown here use $t=1/2$ for the rescaled hopping, leading to a bandwidth of 1 for the non-interacting system. 

\subsection{DMFT on the Bethe lattice}

DMFT~\cite{Georges96} revolves around three dynamical (i.e., frequency-dependent) objects, namely the (interacting) Green's function $G$, the bare Green's function $G_0$ and the self-energy $\Sigma$, which are connected by Dyson's equation,
\begin{align}
    G^{-1}(i\nu_n) = G^{-1}_0(i\nu_n) - \Sigma(i\nu_n). \label{eq:dyson}
\end{align}
Here, $i\nu_n=i(2n+1)\pi T$ denotes the $n$-th Matsubara frequency. Dyson's equation is evidently diagonal in the Matsubara representation: it does not couple different Matsubara frequencies $\nu_n \neq \nu_m$, since the same frequency $\nu_n$ appears in all three objects~\cite{* [{See also the recent discussion in }] Aryasetiawan22}. Paramagnetism is assumed and all spin labels are suppressed. 

The self-consistency condition for the Bethe lattice is 
\begin{align}
    G_0^{-1}(i\nu_n) &= i\nu_n - t^2 G(i\nu_n).\label{eq:bethe}
\end{align}
This relation is also diagonal in the Matsubara frequency. The self-consistency condition can also be expressed in terms of the hybridization function $\Delta$,
\begin{align}
    G_0^{-1}(i\nu_n) &\equiv i\nu_n - \Delta(i\nu_n), \\
    \Delta(i\nu_n) &= t^2 G(i\nu_n).\label{eq:delta:sc}
\end{align}
It should be noted that this linear relationship between $\Delta$ and $G$ is a peculiarity of the Bethe lattice, for a general lattice $\Delta(i\nu_n)$ is a more complicated function of $G(i\nu_n)$.

The final ingredient for DMFT calculations is the solution of the many-body impurity problem defined by $G_0$, using a so-called impurity solver. For the present discussion, this simply means an algorithm to calculate the functional relationship $\Sigma(G_0;U,\beta)$, which can be highly non-linear and couples all frequencies of the input $G_0$ and the output $\Sigma$. 

In total, the DMFT self-consistency cycle consists of three relations $G_0\overset{\text{impurity}}{\mapsto} \Sigma \overset{\text{Dyson}}{\mapsto} G \overset{\text{SC}}{\mapsto} G_0$. The physical parameters $U$ and $t$ appear in the self-energy and the self-consistency condition, respectively. The temperature $T$ only enters implicitly, via the domains of the dynamical fields. 

\subsection{Dynamical fields: representations}

In the imaginary time representation, the Green's function is defined on the domain $[0,\beta)$, i.e., $G(\tau):[0,\beta)\rightarrow \mathbb{R}$, where $\beta=1/T$. The Green's function can then be extended anti-periodically to all $\tau$ via the relation $G(\tau-\beta)=-G(\tau)$, reflecting the fermionic nature of the electron. The Matsubara representation is obtained by performing a Fourier transformation, $G(i\nu_n) = \int_0^\beta d\tau \, G(\tau) \exp(+i\nu_n \tau)$, $G(\tau) = \frac{1}{\beta} \sum_{\nu_n} G(i\nu_n) \exp(-i\nu_n \tau)$.
From the properties of $G(\tau)$, it follows that $G(-i\nu_n)=G(i\nu_n)^\ast$, so it is sufficient to consider the positive frequencies only. 

The imaginary time and Matsubara frequency representations are not particularly \emph{compact}~\cite{Boehnke11}. In imaginary time, a very fine discretization mesh is needed to faithfully describe the continuous function. In the Matsubara representation, $G(i\nu_n)$ decays only algebraically (see also Appendix~\ref{app:tails}) and a large number of frequencies is needed to give a full description of the Green's function. The origin of this slow decay is the discontinuity of $G(\tau)$ and its derivatives at $\tau=0$. Representing $G(\tau)$ in terms of appropriately chosen orthogonal basis sets turns out to be much more efficient. Here, the Legendre polynomials are considered, following the notation of Ref.~\cite{Boehnke11}, $
    G(l) = \sqrt{2l+1} \int_0^\beta d\tau\, G(\tau) P_l(2\tau/\beta-1)$, 
    $G(\tau) = \sum_{l\geq 0} \frac{\sqrt{2l+1}}{\beta} G(l) P_l(2\tau/\beta-1)$.
Note that the coefficients $G(l)$ are real.

The basis transformations to the Matsubara and Legendre representations are unitary and preserve the norm,
\begin{align}
    \norm{G}^2 = \sum_{n=-\infty}^\infty \abs{G(i\nu_n)}^2=\beta \int_0^\beta \abs{G(\tau)}^2 d\tau= \sum_l \abs{G(l)}^2. \label{eq:norm}
\end{align}

Altogether, the dynamical fields $G$, $G_0$ and $\Sigma$ can be seen as vectors in an infinite-dimensional vector space, and the various representations provide different basis sets for this vector space. At the same time, the physical Green's functions form only a small subset (embedded manifold) of this vector space. 

For the Bethe lattice at half-filling, as considered in this work, there is an additional symmetry (particle-hole symmetry), which further restricts the space of physical Green's functions, since $G(\tau) = G(\beta-\tau)$, $G(\tau) = -G(-\tau)$, $\Re G(i\nu_n) = 0$, $G(i\nu_n) =-G(-i\nu_n)$, and $G(l)=0$ for $l$ odd.
Similar properties hold for the bare Green's function $G_0$ and the self-energy $\Sigma$, after removing the Hartree term in the latter.
For the Matsubara representation, this means that it is only necessary to consider the imaginary parts of the Green's function, so $G : \mathbb{N} \rightarrow \mathbb{R}, n\mapsto \Im G(i\nu_n)$ is an element of the real vector space $\mathbb{R}^\mathbb{N}$. This will be convenient when discussing derivatives, which are all real.

\subsection{Free energy functional}
\label{sec:Landaufunctional}

Following Landau's theory of phase transitions, a free energy functional can be used to describe the hysteresis region of DMFT. For the Bethe lattice, this approach is described in detail in Section 3.7 of Ref.~\cite{Blumerphd} and references therein~\cite{Georges92b,Kotliar00}. The lattice free energy functional can be written as $\Omega[\Delta]=\Omega^\text{imp}-\Omega'$. Here $\Omega^\text{imp}$ is the free energy function of the impurity model, which satisfies $\frac{1}{T} \delta \Omega^\text{imp}/\delta \Delta(i\nu)=G(i\nu)$. Functional derivatives $\delta \Delta$ are used for the free energy functional $\Omega$, partial derivatives $\partial \Delta$ for other derivatives. The free energy associated with the lattice, $\Omega'=\frac{T}{2t^2} \sum_\nu \Delta(i\nu)^2$, ensures that the DMFT self-consistency condition is fulfilled at the extremal points of $\Omega$, since
\begin{align}
    0&=\frac{1}{T} \frac{\delta \Omega}{\delta \Delta(i\nu)} = G(i\nu)-\frac{ \Delta(i\nu) }{t^2}
\end{align}
is the self-consistency condition of the Bethe lattice. 

The second derivative of the free energy functional distinguishes stable (minimum) and unstable (maximum) solutions of the self-consistency equations. It is a matrix in Matsubara space,
\begin{align}
    \frac{1}{T} \frac{\delta^2 \Omega}{\delta \Delta(i\nu)\delta \Delta(i\nu')} &= \frac{\partial G(i\nu)}{\partial \Delta(i\nu')} - \frac{\delta_{\nu\nu'}}{t^2} \notag \\
    &= -G^2(i\nu)\frac{\partial G^{-1}(i\nu)}{\partial \Delta(i\nu')} - \frac{\delta_{\nu\nu'}}{t^2} \notag \\
    &= G^2(i\nu) \left[\delta_{\nu\nu'} + \frac{\partial \Sigma(i\nu)}{\partial \Delta(i\nu')}\right] - \frac{\delta_{\nu\nu'}}{t^2} \notag \\ 
    &= G^2(i\nu) \left[\delta_{\nu\nu'} + T  F_{\nu\nu'} G^2(i\nu')\right] - \frac{\delta_{\nu\nu'}}{t^2} \notag \\
    &= \hat{\chi}_{\nu\nu'} - \frac{\delta_{\nu\nu'}}{t^2}. \label{eq:landau:foript}
\end{align}
The basic idea is that the derivative $\partial/\partial \Delta$ introduces an additional $c^\dagger c$ in any expectation value, so $\partial G/\partial \Delta$ is a two-particle correlation function. Accounting for prefactors, following Ref.~\cite{vanLoon20}, and with $\omega=0$ implied everywhere,  
$
\partial \Sigma(i\nu)/\partial \Delta(i\nu') = \frac{1}{\beta} F_{\nu\nu'} G^2(i\nu'),
$
where $F$ is the reducible vertex of the impurity model. Finally, $\hat{\chi}=GG+GGFGG$ is the generalized susceptibility of the impurity model. The relation between $\hat{\chi}$ and $F$ is a local Bethe-Salpeter equation.

These equations are written entirely in terms of the reducible vertex or the susceptibility of the impurity model. Since the impurity model at finite temperature cannot feature any phase transition, these objects are well-defined and divergence free. In particular, the inversion of a local Bethe-Salpeter to obtain the irreducible vertex is avoided in this formulation.

In this Landau theory formulation of DMFT~\cite{Kotliar00,Strand11,vanLoon20}, the Hessian $\delta^2\Omega/\delta\Delta^2$ is positive definite for stable solutions and has (at least) one negative eigenvalue for unstable solutions. At the critical point, one eigenvalue is exactly equal to zero, signalling the transition from a stable to an unstable solution. This also happens for the disappearing solution at the edge of the hysteresis region, where a stable and an unstable solution merge. The distinction between the critical point and the hysteresis region boundary can thus be found in the third derivative of the free energy: at the critical point $\delta^3 \Omega/\delta \Delta^3=0$ along the ``direction'' given by the vanishing second derivative. 

\begin{figure}
 \includegraphics{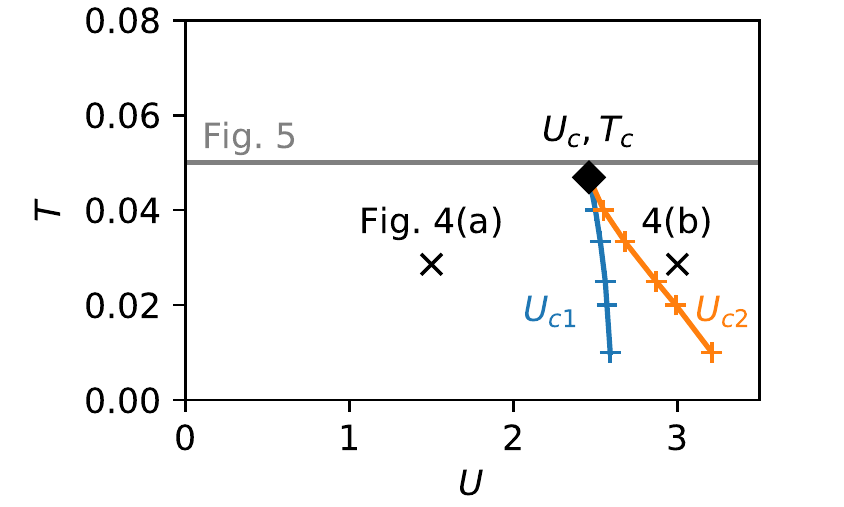}
 \caption{Phase diagram of IPT for the Bethe lattice. The position of the critical point at $U_c\approx 2.46$ and $T_c \approx 4.69 \cdot 10^{-2}$,
i.e., $\beta_c\approx 21.3$ is taken from Ref.~\onlinecite{Strand11}. $U_{c1}$ and $U_{c2}$ denote the boundaries of the hysteresis region, the parameters of Figs.~\ref{fig:U1.5} and \ref{fig:beta20} are also marked. }
 \label{fig:pd}
\end{figure}

For the Bethe lattice, $\Omega'$ is quadratic in $\Delta$, because of the linear form of  Eq.~\eqref{eq:bethe}. This implies $\delta^3 \Omega/\delta \Delta^3=\delta^3 \Omega^\text{imp}/\delta \Delta^3$, since $\Omega'$ is quadratic in $\Delta$. 

\subsection{Iterative properties of DMFT}
\label{sec:iterative}

DMFT is a self-consistent theory, similar to Curie-Weiss mean-field theory for magnetism. 
Although only converged solutions have a formal role in mean-field theory, useful physical insight can be gained by studying the convergence towards such a solution, since the convergence speed is related to the presence of ``soft modes'' and enhanced response~\cite{vanLoon20}. For Curie-Weiss theory, this is illustrated in Appendix~\ref{app:curieweiss}.

Where Curie-Weiss theory considers a scalar as the self-consistently determined mean-field, DMFT uses the infinite-dimensional vector $\Delta(i\nu_n)$ instead. In both cases, forward iteration provides a way to achieve self-consistency. Starting from an initial guess $G_0$, one calculates $G_0 \mapsto \Sigma \mapsto G \mapsto G_0$ until convergence is reached. 

The convergence of forward iteration is determined by the DMFT Jacobian~\cite{Blumerphd,Zitko09,Strand11,vanLoon20}.
The mathematical theory of self-consistent schemes states that the eigenvalues $\lambda_\alpha$ and eigenvectors $v_\alpha$ of the Jacobian $J$ determine the iterative flow of the self-consistent scheme $\Delta^{(n+1)}=f(\Delta^{(n)})$ in the linear regime around a self-consistent solution $\Delta^\ast$. Starting with an initial guess $\Delta^{(0)}$ sufficiently close to $\Delta^\ast$, the difference $\Delta^{(n)}-\Delta^{\ast}$ can be expanded in the eigenbasis of $J$, $\Delta^{(n)}-\Delta^{\ast}=\sum_\alpha c_\alpha v_\alpha$ and the subsequent iterations give
\begin{align}
 \Delta^{(n)}-\Delta^{\ast}\approx \sum_\alpha \lambda_\alpha^n c_\alpha v_\alpha
\end{align}
In the eigenbasis, every component evolves according to its own eigenvalue. The solution converges exponentially as long as $\abs{\lambda_\alpha}< 1$ for all eigenvalues of $J$. Furthermore, smaller (absolute) eigenvalues lead to faster convergence. 

The relevant Jacobian in DMFT is an infinite-dimensional matrix, since the field is infinite-dimensional.  
Note that the Jacobian keeps track of how a single object in the self-consistency cycle changes during the iterative process, so the expression for the Jacobian differs based on the object that is tracked, but the convergence speed turns out to be identical, as shown below.

The Jacobian for $G_0$ is obtained by taking the derivative of the self-consistency condition, 
\begin{align}
    J^{G_0}_{\nu\nu'} 
    &= \frac{\partial G_0^\text{new}(i\nu)}{\partial G_0^\text{old}(i\nu')} \notag \\
    &= -G_0^2(i\nu) \frac{\partial (G^{-1}_0)^\text{new}(i\nu)}{\partial G_0^\text{old}(i\nu')} \notag \\
    &= t^2 G_0^2(i\nu)\frac{\partial G^\text{old}(i\nu)}{\partial G^\text{old}_0(i\nu')} \notag \\
    &= t^2 G^2(i\nu)\left(\delta_{\nu\nu'}+G_0^2(i\nu) \frac{\partial \Sigma(i\nu)}{\partial G_0(i\nu')} \right).\label{eq:jacobian}
\end{align}
Here, in the last equation, the label old is dropped, since it appears on both sides of the derivative.  
Similarly, the Jacobian for the hybridization function $\Delta$ is obtained by taking the derivative of the self-consistency condition $\Delta^\text{new}(i\nu) = t^2 G^\text{old}(i\nu)$
\begin{align}
    J^{\Delta}_{\nu\nu'} &= \frac{\partial \Delta^\text{new}(i\nu)}{\partial \Delta^\text{old}(i\nu')} \notag \\
    &= t^2 \frac{\partial G^\text{old}(i\nu)}{\partial \Delta^\text{old}(i\nu')} \notag \\
    &= t^2 G^2(i\nu) \left(\delta_{\nu\nu'}+\frac{\partial \Sigma(i\nu)}{\partial \Delta(i\nu')}\right) \notag \\
    &= t^2 G^2(i\nu)\left(\delta_{\nu\nu'}+\frac{\partial \Sigma(i\nu)}{\partial G_0(i\nu')} G_0^2(i\nu')\right).\label{eq:jacobian:alt}
\end{align}
The two Jacobians are \emph{similar}, since $J^{\Delta} = Q^{-1} J^{G_0} Q$ for $Q_{\nu\nu'}=\delta_{\nu\nu'} G_0^2(i\nu')$. Thus, they have the same eigenvalues and the forward iteration converges with the same speed, even though the eigenvectors are different.

To get closer to the notation of Ref.~\cite{vanLoon20}, we recall that
$
T F_{\nu\nu'} G^2(i\nu')
    =
    \frac{\partial \Sigma(i\nu)}{\partial \Delta(i\nu')}$. 
This gives
\begin{align}
J^{\Delta}_{\nu\nu'} &= t^2 G^2(i\nu) \delta_{\nu\nu'} + t^2 T \, G^2(i\nu) F_{\nu\nu'} G^2(i\nu')=t^2\hat{\chi}_{\nu\nu'}. \notag
\end{align}

The Jacobian can be split into two terms, with and without a contribution from the vertex,
\begin{align}
    J^\Delta_{\nu\nu'} &\equiv J^0_{\nu\nu'}+J^1_{\nu\nu'},\\
    J^0_{\nu\nu'} &= t^2 G^2(i\nu) \delta_{\nu\nu'} \notag \\
    J^1_{\nu\nu'} &= t^2 T \, G^2(i\nu) F_{\nu\nu'} G^2(i\nu'). \notag
\end{align}

\emph{Relation to free energy functional}: Based on the free energy functional, a more convenient derivation of the Jacobian is possible. $\Delta = t^2 G = t^2 \delta\Omega^\text{imp}/\delta \Delta$ so $J^\Delta=\partial \Delta^\text{new}/\partial \Delta^\text{old}=t^2 \delta^2 \Omega^\text{imp}/\delta \Delta^2=t^2 \hat{\chi}$. This reflects the fact that the self-consistency condition itself is the derivative of the scalar function $\Omega$. In that case, according to Clairaut's theorem, the Jacobian $J^\Delta$ is symmetric. This is equivalent to the symmetry $F_{\nu\nu'}=F_{\nu'\nu}$.

However, as will be discussed in Sec.~\ref{sec:freeenergy:ipt} and \ref{sec:jacobian:ipt}, IPT cannot be derived from a free energy functional, but the derivation of the Jacobian given here remains valid for IPT.

\subsection{Jacobian, lattice free energy and the metal-insulator transition}

The relation between the Jacobian and the Hessian of the impurity free energy also provides a direct relation to the lattice free energy functional,
\begin{align}
    \frac{1}{T} \frac{\delta^2 \Omega}{\delta \Delta(i\nu)\delta \Delta(i\nu')}= \frac{1}{t^2} (J^\Delta_{\nu\nu'} - \delta_{\nu\nu'}).
\end{align}
Thus, the leading eigenvalue of the Jacobian, $\lambda_J$, is closely related to the metal-insulator phase diagram~\cite{Blumerphd,Strand11,vanLoon20}, for IPT the phase diagram is shown in Fig.~\ref{fig:pd}. Above $T_c$, all eigenvalues of the Jacobian are smaller than unity in absolute value, i.e., $\abs{\lambda_J}<1$. The resulting solution is a minimum of the free energy functional~\footnote{Note that $\Delta(i\nu)$ is purely imaginary, so the sign of $\delta^2\Omega/\delta(i\Delta)^2$ should be considered, which results in a minus sign.}. Exactly at $(U_c,T_c)$, one eigenvalue is equal to unity. In the hysteresis region at $T<T_c$, the metallic and insulating solutions have eigenvalues smaller than unity, so they are stable. There is a third, unstable solution with eigenvalue larger than unity. At the boundary of the hysteresis region, the leading eigenvalues of the disappearing solutions (one stable and one unstable) are both equal to unity. Together, the leading eigenvalues of the possible solutions form a smooth curve, which implies an infinite slope of the disappearing solutions at the edge of the hysteresis region. This situation is illustrated in Fig. 12 of Ref.~\cite{Strand11}. At $T_c$, the leading eigenvalue reaches unity at $U_c$, and then goes back down. 

\section{Iterated Perturbation Theory}
\label{sec:ipt}

The missing ingredient in the discussion so far is the functional relation $\Sigma[G_0]$. In IPT, a particularly simple approximation for $\Sigma[G_0]$ is used, namely
\begin{align}
    \Sigma^\text{IPT}[G_0] = -U^2 G_0(\tau)G_0(-\tau) G_0(\tau). \label{eq:ipt}
\end{align}
This expression corresponds to applying second order perturbation theory (SOPT, also called GF2) as the impurity solver, under the assumption of particle-hole symmetry. The distinction between second-order perturbation theory and IPT is the way that the ``input'' $G_0$ is determined. In IPT, it is determined self-consistently according to the DMFT self-consistency condition. In second-order pertubation theory, on the other hand, it is either fixed to the initial $G_0$ or it is updated according to the Dyson equation in self-consistent SOPT. In the end, this difference is responsible for the fact that IPT has a metal-insulator transition and SOPT does not.

At particle-hole symmetry, the self-energy simplifies to $\Sigma^\text{IPT}(\tau)=U^2 G_0(\tau)^3$. 

Note that this expression for the IPT self-energy ignores the Hartree term, which is of first order in $U$ and is instantaneous. When the density is fixed at half-filling, the Hartree term cancels with the chemical potential and there is no need to keep track of it explicitly.

\subsection{The IPT vertex}
\label{sec:vertex}

\begin{figure}
\begin{tikzpicture}

\node at (-0.7,0) {$\Sigma=$} ;
\node[fill=black,circle] (left) at (0,0) {};
\node[fill=black,circle] (right) at (1.5,0) {};

\draw[->,out=45,in=90+45] (left) to node[above] {$G_0$} (right) ;
\draw[->,out=-45,in=-90+-45] (left) to (right) ;
\draw[<-,out=0,in=180] (left) to (right) ;

\end{tikzpicture}\\
\begin{tikzpicture}

\node at (-0.9,0) {$\dfrac{\partial \Sigma}{\partial G_0}=$} ;

\node[fill=black,circle] (left1) at (0,0) {};
\node[fill=black,circle] (right1) at (0.9,0) {};
\draw[->,out=45,in=90+45] (left1) to  (right1) ;
\draw[->,out=-45,in=-90+-45] (left1) to (right1) ;

\node[fill=black,circle] (left2) at (1.8,0) {};
\node[fill=black,circle] (right2) at (2.7,0) {};
\draw[->,out=45,in=90+45] (left2) to  (right2) ;
\draw[<-,out=0,in=180] (left2) to (right2) ;

\node[fill=black,circle] (left3) at (3.6,0) {};
\node[fill=black,circle] (right3) at (4.5,0) {};
\draw[->,out=-45,in=-90+-45] (left3) to (right3) ;
\draw[<-,out=0,in=180] (left3) to (right3) ;

\node at (1.35,0) {$+$} ;
\node at (1.8+1.35,0) {$+$} ;
\end{tikzpicture}
    \caption{Iterated Perturbation Theory self-energy and vertex}
    \label{fig:iptdiagrams}
\end{figure}
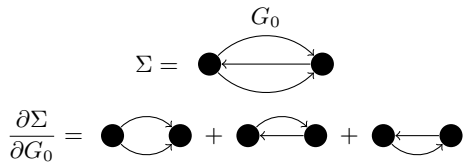

To determine the stability of DMFT, we need the derivative of the self-energy of the impurity model with respect to the input, i.e., $\partial \Sigma/\partial G_0$ or $\partial \Sigma/\partial \Delta$ should be calculated. The relation between the two derivatives is
\begin{align}
    \frac{\partial \Sigma(i\nu)}{\partial \Delta(i\nu')} &= \frac{\partial \Sigma(i\nu)}{\partial G_0(i\nu')} G_0^2(i\nu').
\end{align}
In this manuscript, all three objects, $F$, $\partial \Sigma/\partial G_0$ and $\partial \Sigma/\partial \Delta$ will be called ``vertex'', the explicit mathematical expression will be used when it is necessary to be specific. 

From $\Sigma(\tau) = -U^2 G_0(\tau)G_0(-\tau) G_0(\tau)$, the derivative $\partial \Sigma/\partial G_0$ can be taken directly,
\begin{align}
    \frac{\partial \Sigma(\tau)}{\partial G_0(\tau')} =& -2U^2 G_0(\tau)G_0(-\tau) \delta(\tau-\tau') \notag \\
    &- U^2 G_0(\tau)^2 \delta(\tau+\tau'), \label{eq:vertex:tau}
\end{align}
or, at particle-hole symmetry, 
\begin{align}
    \left(\frac{\partial \Sigma(\tau)}{\partial G_0(\tau')}\right)^\text{PHS} = 3U^2 G^2_0(\tau) \delta(\tau-\tau').
\end{align}

Diagrammatically, taking a derivative corresponds to cutting a single line from every possible self-energy diagram in every possible way, see Fig.~\ref{fig:iptdiagrams}. The IPT self-energy functional consists of a single diagram with three internal lines $G_0$, so there are three possible ways to cut a line, giving the prefactor 3. 

The derivative, $\frac{\partial \Sigma(\tau)}{\partial G_0(\tau')}$, interpreted as a matrix in imaginary time, is diagonal with strictly positive elements on the diagonal. The operator is therefore positive definite if $U\neq 0$, which is a property that is independent of the choice of basis, i.e., it also holds in the Matsubara basis. The matrix also has a well-defined inverse (again, if $U\neq 0$), signalling that the IPT-relation between $\Sigma$ and $G_0$ is invertible. As a diagonal matrix, $\frac{\partial \Sigma(\tau)}{\partial G_0(\tau')}$ is also symmetric. This is actually an undesirable property of IPT, since it implies that $F_{\nu\nu'}$ is not symmetric within the IPT approximation. 

It is also useful to express the self-energy and the functional derivative in the Matsubara representation. This can be done diagrammatically, or by Fourier transforming the previously obtained imaginary time expressions, see also Appendix~\ref{app:vertexreps}. Either way, 
\begin{align}
    \Sigma(i\nu) = -\frac{U^2}{\beta^2} \sum_{\nu_1\nu_2} G_0(i\nu_1) G_0(i\nu_2) G_0(i\nu_1+i\nu_2-i\nu), \label{eq:sigma:matsubara}
\end{align}
leading to
\begin{align}
    \frac{\delta \Sigma(i\nu)}{\delta G_0(i\nu')} &= -2\frac{U^2}{\beta^2} \sum_{\nu_1} G_0(i\nu_1)G_0(i\nu_1+i\nu'-i\nu) \notag \\ &\phantom{=} - \frac{U^2}{\beta^2} \sum_{\nu_1} G_0(i\nu_1)G_0(i\nu+i\nu'-i\nu_1) 
\end{align}
The functional derivative $\delta \Sigma(i\nu)/\delta G_0(i\nu')$ is a matrix in $\nu,\nu'$. Since there is a relation between the value of the Green's function at positive and negative Matsubara frequencies, it is generally sufficient to consider the symmetric and antisymmetric parts of the matrix separately~\cite{Springer20}. 
At particle hole-symmetry, only the functional derivative within the particle-hole symmetric manifold ($G_0(i\nu)=-G_0(-i\nu)$) is needed~\footnote{A related question is if the $G_0$ is physical, in the sense that it is the Laplace transform of a positive semi-definite density of states, and if the functional derivative can be constrained to the manifold of physical Green's functions. Since the IPT expression for the self-energy is written entirely in imaginary time, however, this question is not our concern here.}. This restricts variations to the form 
\begin{align}
    \delta G_0(\abs{\nu})\equiv \delta G_0(i\nu) - \delta G_0(-i\nu). \label{eq:deltaG:antisymmetric}
\end{align}
The associated functional derivative $\delta \Sigma(\nu)/\delta G({\abs{\nu'}})$ is
\begin{align}
    \delta \Sigma(\nu)/\delta G_0(\abs{\nu'}) = -3 \frac{U^2}{\beta^2} \sum_{\nu_i} &\left[ G_0(i\nu_i)G_0(i\nu_i+i\nu'-i\nu) \right. \notag\\ &\left.-G_0(i\nu_i)G_0(i\nu_i-i\nu'-i\nu) \right]. \label{eq:dSigmadG0:antisymmetrized}
\end{align}
More details on the structure of the matrix $\partial \Sigma/\partial G_0$ are available in Appendix~\ref{app:vertexreps}.

Having determined the vertex $\partial \Sigma/\partial G_0$, it is also possible to consider higher-order derivatives. In IPT, $\partial^2 \Sigma/\partial G_0^2$ and $\partial^3 \Sigma/\partial G_0^3$ are finite, and all higher orders are equal to zero. This follows from the explicit IPT form of the self-energy and is a notable difference with the exact solution.

\subsection{IPT's problems on the two-particle level: Non-existence of free energy functional}
\label{sec:freeenergy:ipt}

The derivation of the free energy functional in Sec.~\ref{sec:Landaufunctional} holds for the general formulation of DMFT. However, a problem is encountered when IPT is used as the impurity solver. As discussed above, $F_{\nu\nu'}$ should be a symmetric matrix, but this is not the case in IPT. This would imply that the Hessian of the free energy functional is not symmetric, which is impossible. Thus, the conclusion is that there is no consistent way to derive IPT from a free energy functional. However, the Jacobian is still useful to characterize the hysteresis region, as will be discussed in Sec.~\ref{sec:jacobian:ipt}.

The lack of a free energy functional might come as a surprise. It is useful to consider an analogy in the form of electromagnetism. In electrostatics, the electric field $\vec{E}$ is curl-free, $\nabla\times \vec{E}=0$, i.e., the matrix $\partial E_i/\partial x_j$ is symmetric. This property guarantees that the electric field can be written as the gradient of a potential $V$, $\vec{E}=-\nabla V$. Given $\vec{E}$, $V(\vec{x})$ can be determined up to a constant by simply taking a line integral of $\vec{E}$ from any point $\vec{x}_0$. Due to the vanishing curl, this construction is independent of the integration path. Now, imagine that we had an approximate way to calculate the electric field for a system of interest, $\vec{E}^\text{approx}(\vec{x})$. If this approximation satisfies $\nabla \times \vec{E}^\text{approx}=0$, then there is still a consistent way to construct $V^\text{approx}$ via line integrals. However, if $\nabla \times \vec{E}^\text{approx} \neq 0$, the construction breaks down. $\vec{E}^\text{approx}$ can still be a useful approximation for the electric field, it just does not allow a discussion of potentials. The situation is similar in IPT, in the sense that the explicit recipe $G(\Delta)$ of the IPT impurity solver is inconsistent with the existence of a free energy functional $\Omega^\text{imp}[\Delta]$ with $G=T^{-1} \partial \Omega^\text{imp}/\partial \Delta$. 

This raises the question why $F$ is not symmetric in IPT, which brings us back to the functional derivative. There is a difference between $\partial/\partial G_0$ and $\partial/\partial \Delta$ in the precise way the cutting of lines in the diagram is handled: in $\partial/\partial G_0$ the ``stumps'' left by the cut are amputated with $G_0$, whereas in $\partial/\partial \Delta$, a factor $G_0$ is attached at the stumps. Finally, in the usual definition of $F$, the ends of the $\partial/\partial \Delta$ expression are divided by $G$. In total, the definition of $F$ amputates all one-particle irreducible contributions from the legs of the vertex. In the exact solution, these are generated by higher orders in $U$ in the expression for $\Sigma$, but in IPT they are absent completely. Thus, in IPT, the definition of $F$ amputates too much. But this problem only occurs in the two legs that are cut when the derivative is taken, the two external legs in the initial expression for $\Sigma^\text{IPT}$ are properly amputated by construction. At a very fundamental level, IPT breaks the equivalence of the four external legs of $F$, and the origin of this equivalence breaking is the use of $G_0$ instead of $G$ in the diagrammatic expression for $\Sigma$~\cite{Baym62}. Thunstr\"om et al.~\cite{Thunstrom18} identified this phenomenon with time-reversal symmetry breaking, but from the diagrammatic analysis, the inequivalence of the legs appears as the origin.

\subsection{Jacobian in IPT}
\label{sec:jacobian:ipt}

The Jacobian for DMFT is given in Equation~\eqref{eq:jacobian}. For an exact solution of the impurity model, $F$ is a symmetric matrix and $J^\Delta$ is thus also symmetric. However, within the IPT approximation, $J^\Delta$ is not symmetric, it is only \emph{similar} to a symmetric matrix: defining $X_{\nu\nu'}=\delta_{\nu\nu'} G(i\nu)/G_0(i\nu)$, the matrix $X^{-1} J^\Delta X$ is real and symmetric. Essentially, one factor $G(i\nu)/G_0(i\nu)$ is divided out from the left in Eq.~\eqref{eq:jacobian:alt} and replaced by a factor $G(i\nu')/G_0(i\nu')$ from the right. Due to the similarity, the non-symmetric $J^\Delta$ still has real eigenvalues and the symmetrized version of $J$ has an orthogonal basis of eigenvectors. The factors $(G/G_0)$ are related to the amputation procedure of the vertex discussed above.

It is useful to analyse several limiting cases of the Jacobian, namely $U=0$, $t=0$ and large $\nu$, since the Jacobian matrix turns out to be diagonal (in the Matsubara representation) in these situations. Additional $T=0$ expressions are given in Appendix~\ref{app:T0}.

\emph{Non-interacting model:} The DMFT self-consistency cycle becomes trivial in the non-interacting model, $U=0$, since it implies $\Sigma=0$, $G_0=G$ and the self-consistency condition becomes $G^{-1}(i\nu_n) = i\nu_n - t^2 G(i\nu_n)$, which is solved by $G(i\nu_n) = \frac{i\nu_n }{2t^2}\left(1-\sqrt{1+4t^2/\nu_n^2 }\right) $.

Since there is no self-energy at $U=0$, the Jacobian is given by the first, diagonal term, i.e, $J_{\nu\nu'}=J^0_{\nu\nu'}=t^2 G_\nu^2 \delta_{\nu\nu'}$. The eigenvalues are  $t^2 G(i\nu_n)^2=-\frac{\nu_n^2}{4t^2} \left(1-\sqrt{1+4t^2/\nu_n^2 }\right)^2$, which takes values in $(-1,0)$ for $\nu_n \neq 0$. So for $T>0$, this solution is iteratively stable, although it becomes unstable at $T=0$. The leading eigenvector is entirely localized on the lowest Matsubara frequency. These results at $U=0$ do not use the IPT approximation and are generally valid.

\emph{Atomic limit:} For $t=0$, the self-consistency condition is $G_0=1/i\nu_n$ and this makes it possible to evaluate the vertex explicitly, as done in Appendix~\ref{app:vertexreps}. The result is that the IPT vertex is proportional to the identity matrix. At $t=0$, the Jacobian is strictly equal to zero. However, for $0<t\ll U$, the atomic $G_0$ and vertex can still be used, while the Jacobian is given only by $J^\Sigma$ and is diagonal. Due to the factors $G^2$ and $G^2_0$, it is decaying as a function of frequency. Importantly, it results in $J>0$ in the atomic limit, since $G^2 G_0^2$ is positive. 

\emph{Large frequency:} In the limit of large frequency, i.e., $\nu,\nu'\ll t,U$, only the term depending on $\nu-\nu'$ in the vertex remains finite. In the Jacobian, this contribution coming from the vertex has to be compared against the $\delta_{\nu\nu'}$ term. The (constant) vertex contribution has an additional prefactor $G_0(i\nu)^2$, which decays at large frequency. Thus, the Jacobian becomes dominated by the diagonal, non-interacting contribution, $J\approx J^0=t^2 G^2(i\nu)\delta_{\nu\nu'}$ at large frequency. The resulting eigenvalues are simply $t^2 G^2(i\nu_n)<0$ and decay asymptotically as $-t^2\nu_n^{-2}$. In other words, as long as $U$ and $t$ are finite, the spectrum of the Jacobian contains a countably infinite number of eigenvalues that approaches 0 from below, and the corresponding eigenvectors are (approximately) localized on a single pair of Matsubara frequencies $\pm i\nu_n$.

\emph{From weak to strong coupling:} The non-interacting and atomic model both have a diagonal Jacobian, but the eigenvalues of the Jacobian change from negative (at $U=0$) to positive (at $U\gg t$). Thus, with increasing $U$, the eigenvalues of the Jacobian cross zero one by one. Since $J^\Delta = t^2 \hat{\chi}$, this is equivalent to zero crossings of the eigenvalues of the susceptibility, responsible for the divergence of the irreducible vertex~\cite{Schafer13,Gunnarsson17,Chalupa18,Springer20}, since it is based on $\hat{\chi}^{-1}$. 

It should be stressed that the zero-crossings of the eigenvalues of the Jacobian are well-defined in the IPT approximation, even though the free energy and the vertices are problematic, as discussed in Sec.~\ref{sec:freeenergy:ipt}.

\begin{figure}
    \includegraphics{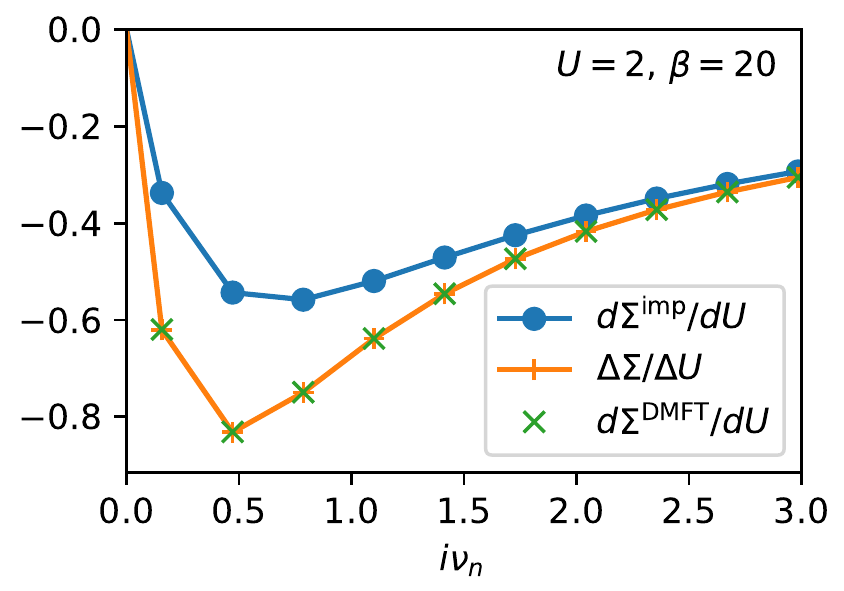}
    \caption{$d\Sigma/dU$ evaluated analytical via Eq.~\eqref{eq:dsigmadU} (green crosses) and numerical via the finite difference $\Delta U=0.001$ (orange plusses). The DMFT response is enhanced compared to the impurity response with $G_0$ held constant (blue circles) due to the self-consistent feedback. }
    \label{fig:dSigmadU}
\end{figure}

\section{Response function: self-energy versus interaction}
\label{sec:response}

The Jacobian also plays a role in the response of the observables of the system to changes in the model parameters. Simply stated, the DMFT response consists of the direct response of the impurity model combined with the self-consistent adjustment of the impurity model, which is described by the Jacobian. The first part is free from thermodynamic divergences by construction (at $T>0$), whereas the second part can have non-smooth behavior when $\lambda_J=1$. As an example, the compressibility has been studied in this way~\cite{vanLoon20}, although it remains smooth in the particle-hole symmetric model~\cite{Eckstein07,Reitner20,vanLoon20} because of the peculiar frequency structure of the leading eigenvector of the Jacobian. This motivates the study of a dynamic quantity, such as the self-energy $\Sigma$. Here, $d\Sigma/dU$ will be considered. 

$U$ appears explicitly only in the self-energy of the impurity model, leading to the derivative $\partial \Sigma^\text{imp}/\partial U$. Here, the label $^\text{imp}$ signals that the derivative is taken while keeping $G_0$ constant.
However, in the DMFT self-consistent procedure, $G_0$ will also vary when $U$ is changed, leading to further changes in the self-energy. The label $^\text{DMFT}$ is used below to denote derivatives where the change in $G_0$ is taken into account. 

The Dyson equation and Eq.~\eqref{eq:bethe} have to be satisfied at all values of $U$, thereby providing relations between the $d/dU$ derivatives of $G_0$, $G$ and $\Sigma$,
\begin{align}
    G^{-2} \frac{dG}{dU} &= G_0^{-2} \frac{dG_0}{dU} + \frac{d\Sigma}{dU}, \notag \\
    \frac{dG_0}{dU} &= G_0^2 t^2 \frac{dG}{dU}.
\end{align}
Note that these relations are all diagonal in the Matsubara representation and Matsubara frequencies are implied as the argument on all objects. Taken together, this gives $(1-t^2 G^2) dG_0/dU = G_0^2 t^2 G^2 d\Sigma/dU$. This enables the calculation of the derivative of the self-energy as
\begin{widetext}
\begin{align}
    \frac{d \Sigma^\text{DMFT}(i\nu_1)}{dU} &=     
    \frac{\partial \Sigma^\text{imp}(i\nu_1)}{\partial U} + \sum_{\nu_2} \frac{\partial \Sigma^\text{imp}(i\nu_1)}{\partial G_0(i\nu_2)} \frac{d G_0(i\nu_2)}{d U}  \notag \\
    &=
    \frac{\partial \Sigma^\text{imp}(i\nu_1)}{\partial U} + \sum_{\nu_2}
    \frac{\partial \Sigma^\text{imp}(i\nu_1)}{\partial G_0(i\nu_2)} \frac{t^2 G_0^2(i\nu_2) G^2(i\nu_2)}{1-t^2 G^2(i\nu_2)} \frac{d \Sigma^\text{DMFT}(i\nu_2)}{d U} , \notag \\    
    \left(\frac{d \Sigma^\text{DMFT}}{dU}\right)_\alpha &=  \left(\frac{d \Sigma^\text{imp}}{dU}\right)_\alpha   + \sum_\beta \hat{C}_{\alpha\beta} \left(\frac{d \Sigma^\text{DMFT}}{dU}\right)_\beta , \notag
    \\
    \frac{d \Sigma^\text{DMFT}}{dU} &= \left(\hat{1}-\hat{C} \right)^{-1} \frac{d \Sigma^\text{imp}}{dU} \label{eq:dsigmadU} , \\
    \hat{C}_{\alpha\beta} &\equiv \frac{\partial \Sigma^\text{imp}(i\nu_\alpha)}{\partial G_0(i\nu_\beta)} \frac{t^2 G_0^2(i\nu_\beta) G^2(i\nu_\beta)}{1-t^2 G^2(i\nu_\beta)} . \notag
\end{align}
\end{widetext}
A numerical illustration of Eq.~\eqref{eq:dsigmadU} is given in Fig.~\ref{fig:dSigmadU}.
After some matrix manipulation, this result can be related to the Jacobian in Eq.~\eqref{eq:jacobian}, 
\begin{align}
 \hat{A}_{\alpha\beta} &\equiv \left[1-t^2 G^2(i\nu_a)\right]_{\alpha\beta} \delta_{\alpha\beta} \notag ,\\
 \hat{P}_{\alpha\beta} &\equiv G_0^2(i\nu_\alpha) G^2(i\nu_\alpha) \delta_{\alpha\beta} \notag ,\\
 \left[\hat{A} - \hat{C}\hat{A}\right]_{\alpha\beta} &= \left[1-t^2 G^2(i\nu_a)\right] \delta_{\alpha\beta} \notag \\
 &\phantom{=}- \left(\frac{\partial \Sigma^\text{imp}}{\partial G_0}\right)_{\alpha\beta} t^2 G_0^2(i\nu_\beta) G^2(i\nu_\beta) \notag \\
 &= \left[1 - \hat{P}^{-1} \hat{J}^{G_0} \hat{P}\right]_{\alpha\beta} ,\notag \\
 \left[\hat{1} - \hat{C}\right] &= \left[1 - \hat{P}^{-1} \hat{J}^{G_0} \hat{P}\right] \hat{A}^{-1}.
\end{align}
The matrices $\hat{A}$ and $\hat{P}$ are diagonal and positive definite at particle-hole symmetry, since $G$ and $G_0$ are purely imaginary. For stable solutions, the leading eigenvalue of the Jacobian $J$ is smaller than unity, which implies that $\hat{1}-\hat{C}$ has only positive eigenvalues. Thus, for a stable solution, Eq.~\eqref{eq:dsigmadU} states that the vector $d\Sigma^\text{DMFT}/dU$ is related to the vector $d\Sigma^\text{imp}/dU$ via the positive definite matrix $\left(\hat{1}-\hat{C}\right)^{-1}$. 

Positive-definiteness is preserved under a basis transformation to imaginary time, so $d\Sigma^\text{DMFT}(\tau)/dU = \int d\tau' \hat{M}_{\tau\tau'} \,\, d\Sigma^\text{imp}(\tau')/dU$, where the positive definite $\hat{M}$ is the Fourier transformed version of $(\hat{1}-\hat{C})^{-1}$. For iterated perturbation theory, $\Sigma^\text{imp}(\tau)=U^2 G_0(\tau)^3$ in Eq.~\eqref{eq:dsigmadU}, and $\partial \Sigma^\text{imp}(\tau)/\partial U = 2U G_0(\tau)^3$ is negative. Combined with the positive-definiteness of the matrix $\hat{M}$, this implies monotonicity of the IPT self-energy for stable solutions at $U>0$:
\begin{align}
\frac{d \Sigma^\text{IPT}(\tau)}{d U}<0 \text{ (monotonicity)}. \label{eq:monotonic}
\end{align}
If one also wishes to consider attractive Hubbard interaction, $U<0$, this statement can be reformulated as $d \Sigma^\text{IPT}(\tau)/d (U^2)<0$, for all $\tau \in [0,\beta)$. Note also that, since $\Sigma(\tau)<0$, there is the implication $d\norm{\Sigma}^2/d(U^2) >0$, which is representation independent, but also a substantially weaker statement. There is no guarantee of monotonicity for individual components of $d\Sigma/dU$ in other representations, this will be illustrated numerically for the Legendre representation.

The proof of monotonicity only holds for stable solutions, $\lambda_J<1$, in other words, it does not apply to the disappearing solutions at the edge of the hysteresis region or at the critical point, where $\lambda_J=1$ and $\partial_U \Sigma$ is divergent, nor to the third, thermodynamically unstable solution that exists within the hysteresis region with $\lambda_J>1$. Exactly at the critical point, $\lambda_J=1$ and the self-energy has infinite slope. Thus, the self-energy shows the expected non-analytical behavior at the critical point, and this derivation shows that the non-analytical behavior is directly related to the iterative stability. It should be noted that not all observables are non-analytical in $U$, since the compressibility has be shown to be smooth at the critical point of the particle-hole symmetric metal-insulator transition~\cite{Reitner20,vanLoon20}.

\begin{figure*}[!]
\includegraphics[]{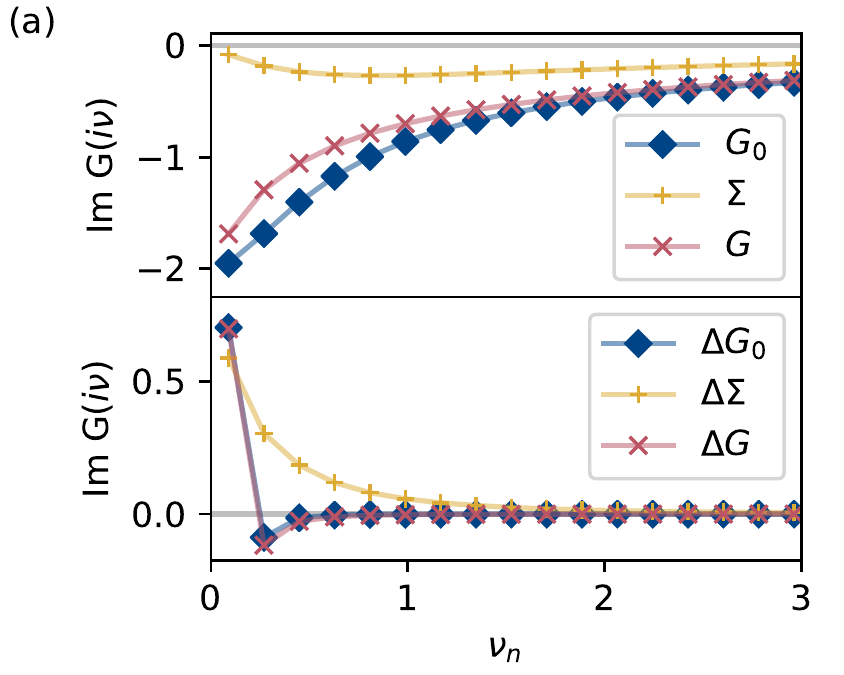}
\includegraphics[]{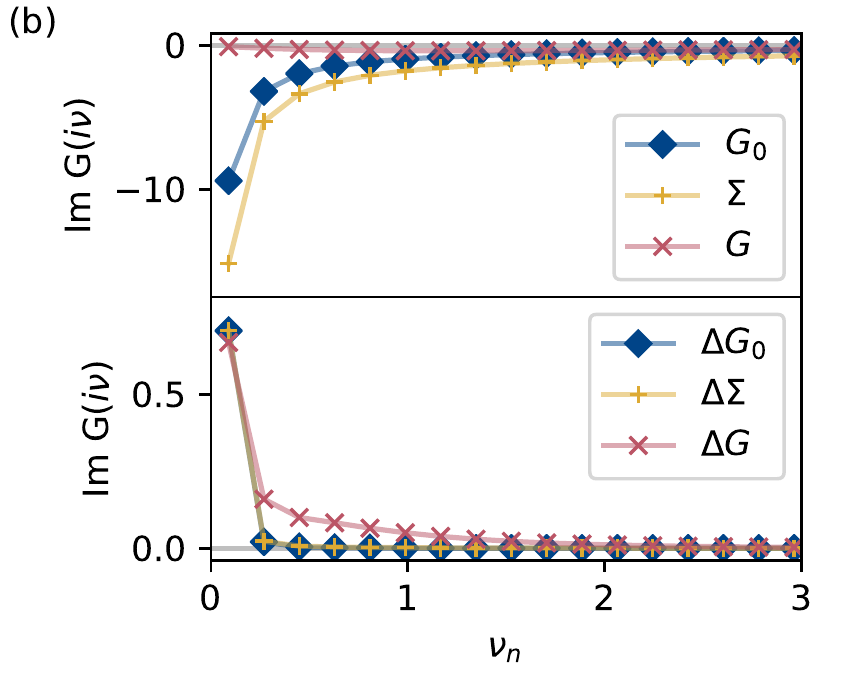}
\includegraphics[]{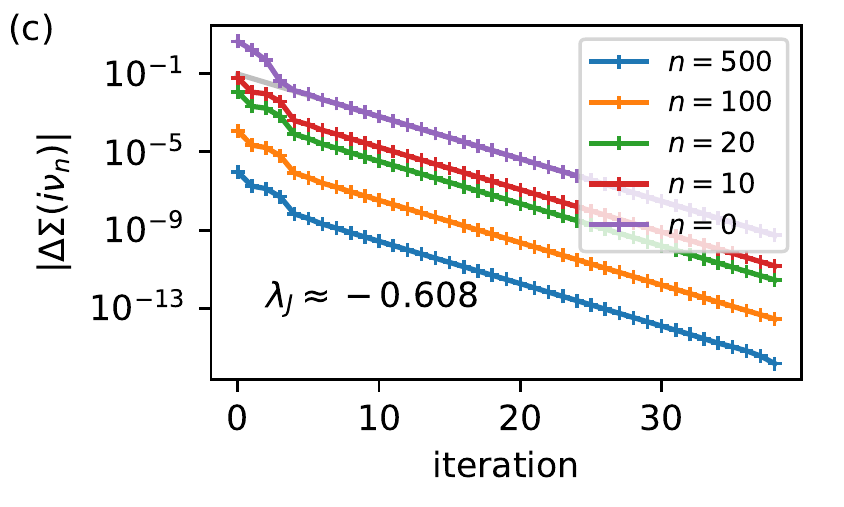}
\includegraphics[]{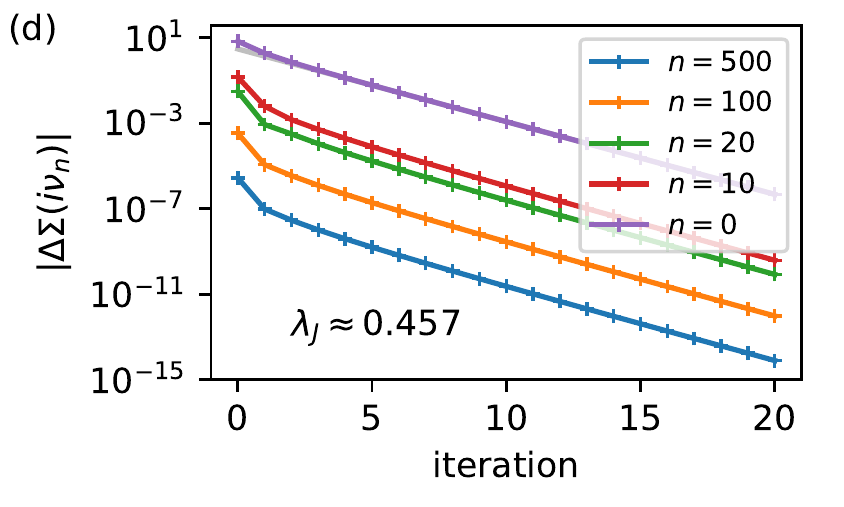}
\includegraphics[]{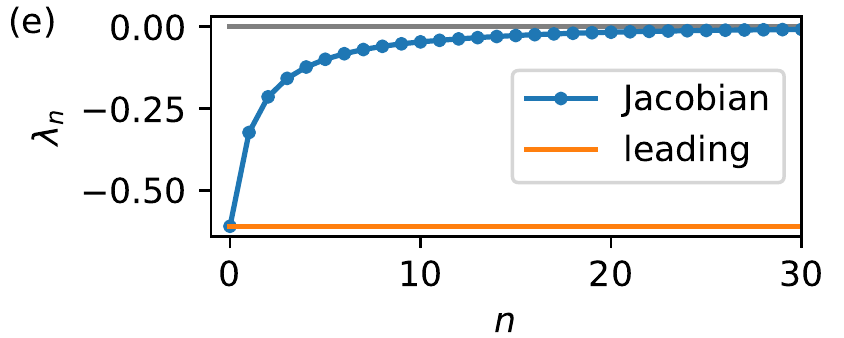}
\includegraphics[]{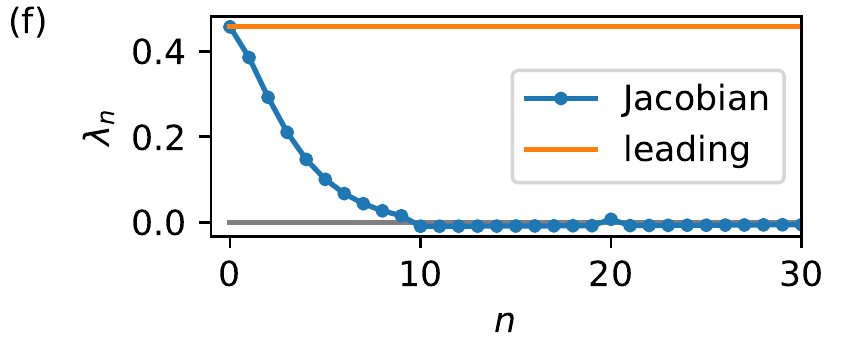}
\caption{Calculations at $\beta=35$, $U=1.5$ (left) and $U=3$ (right). (a-b) converged solution and normalized difference vector of $G_0$, $\Sigma$ and $G$. (c-d) the self-energy converges exponentially in the forward iteration, $\Delta \Sigma(i\nu_n) \propto \lambda_J^\text{iteration}$ at all frequencies. (e-f) The first eigenvalues of the Jacobian, sorted by their absolute value. The value of $\lambda_J$ found in (c-d) corresponds to the leading  eigenvalue here.}
\label{fig:U1.5}
\end{figure*}

\section{Numerical results}
\label{sec:numerics}

Iterated Perturbation Theory is very suitable for numerical studies, since it is fast and does not suffer from stochastic noise. This is especially useful for the derivatives considered in this work, since IPT is sufficiently stable for finite-difference evaluations. Here, a modified version of the IPT code available in TRIQS~\cite{TRIQS} is used to perform calculations, the code is available at \cite{code}. In high-accuracy computations, it becomes necessary to enforce symmetry properties in the calculation explicitly, to avoid minor round-off errors from causing problems. 

Figure~\ref{fig:U1.5} shows the results of calculations at $U=1.5$ (metal) and $U=3$ (insulator), both for $\beta=35$. In Fig.~\ref{fig:U1.5}(a), the metallicity is visible in the linearly vanishing $\Sigma$ around $\nu_n=0$ while Fig.~\ref{fig:U1.5}(b) has a divergent self-energy characteristic of a Mott insulator. Also shown in Fig.~\ref{fig:U1.5}(a-b) are the difference vectors, which are the leading eigenvectors of the respective Jacobians. In Fig.~\ref{fig:U1.5}(a), there is a clear difference in frequency structure between the difference vectors $\Delta G_0$ and $\Delta G$ on the one hand, which feature a sign change between the lowest Matsubara frequency and the higher Matsubara frequencies, and $\Delta \Sigma$ which has a continuous decay with a uniform sign. Although the three \emph{similar} Jacobians have the same eigenvalues, they have different eigenvectors, which is reflected here. 

For $U=3$, in Fig.~\ref{fig:U1.5}(b), the shape of the difference vectors has changed substantially. Now, the difference vectors $\Delta \Sigma$ and $\Delta G_0$ are very similar, whereas $\Delta G$ differs. Furthermore, a sign change within the difference vector no longer occurs. The difference vectors for $G_0$ and $\Sigma$ are almost entirely localized on the lowest Matsubara frequency, as would be expected close to the atomic limit. These changes reflect the growing important of the vertex contribution to the Hamiltonian.

\begin{figure}
    \centering
    \includegraphics{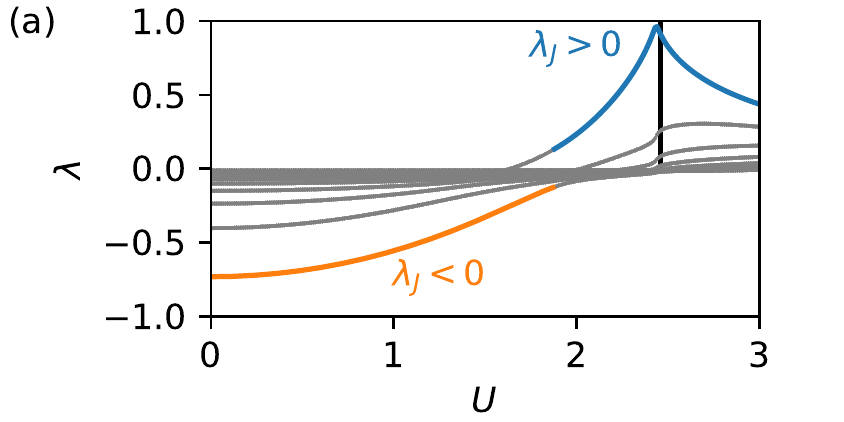}
    \includegraphics{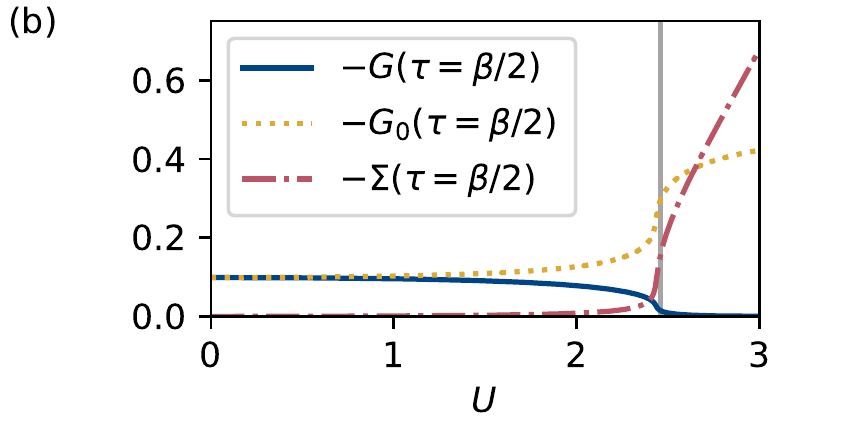}
    \includegraphics{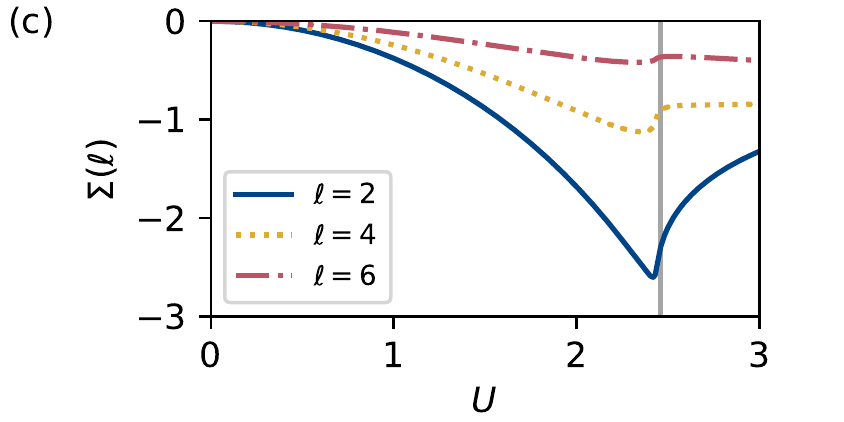}
    \caption{Results at $\beta=20$, which is just above the hysteresis region. (a) Eigenvalues of the Jacobian. A thick colored line is used to highlight the leading eigenvalue, i.e., the one with the largest absolute value. For small $U$, the leading eigenvalue is negative, while all eigenvalues eventually turn positive in the atomic limit. 
    The leading eigenvalue reaches a maximum just below unity close to $U_c$ (vertical line, value from Strand et al.~\cite{Strand11}).   (b) The imaginary-time Green's functions and self-energy evaluated at $\tau=\beta/2$. They change rapidly in the region where $\lambda_J\approx +1$. (c) The evolution of $\Sigma$ in the Legendre basis shows that not all properties are monotonous functions of $U$.}
    \label{fig:beta20}
\end{figure}

Figure~\ref{fig:U1.5}(c-d) shows the convergence towards the self-consistent solution. For both values of $U$, all frequencies are eventually exponentially converging with a single exponent, confirming that there is a single relevant eigenvalue of the Jacobian and that the corresponding eigenvector is finite at all Matsubara frequencies. In fact, $\Delta \Sigma(i\nu) \sim \nu^{-3}$ at large frequencies, since only odd powers in the asymptotic expansion are allowed at particle-hole symmetry, and the $\nu^{-1}$ coefficient of $\Sigma$ is fixed by the parameter $U$, see Appendix~\ref{app:tails}. The simplicity of IPT makes it possible to track the convergence with high accuracy, as is clear from the values on the y-axis.

Finally, Fig.~\ref{fig:U1.5}(e-f) shows the eigenvalues of the Jacobian evaluated according to Eq.~\eqref{eq:jacobian} as blue circles, with the orange line giving the scaling obtained directly from the iterations in Fig.~\ref{fig:U1.5}(c-d). This confirms that the largest absolute eigenvalue determines the convergence. For $U=1.5$, Fig.~\ref{fig:U1.5}(e), all eigenvalues are negative, as in the limit $U=0$. The second eigenvalue is already almost half as small as the leading eigenvalue. For $U=3$, Fig.~\ref{fig:U1.5}(f), the leading eigenvalue of the Jacobian has changed sign, as anticipated by the results for the atomic limit. At the same time, a large number of negative eigenvalues remains (note that the eigenvalues are sorted by their absolute value), as expected from the high-frequency limit.

To illustrate the approach to the critical point, Fig.~\ref{fig:beta20} shows a scan over $U$ at $\beta=20$, i.e., just above $T_c=1/21.3$. The eigenvalues in Fig.~\ref{fig:beta20}(a) start out negative at $U=0$, as expected for the non-interacting limit. This continues until $U\approx 1.64$, where a positive eigenvalue appears, and from $U\approx 1.9$ the leading eigenvalue is positive. This is a sign that $J^\Sigma$ becomes more important than $J^0$. Every eigenvalue that crosses zero corresponds to a divergence in the irreducible vertex~\cite{Schafer13}, and it is clear that these divergences occur before $U_c$. This is a sign that qualitative changes due to correlations effects set in long before the metal-insulator transition is reached~\cite{Schafer13,Springer20}.

As anticipated in Sec.~\ref{sec:response}, strongly enhanced response takes place when $\lambda_J\approx 1$, see Fig.~\ref{fig:beta20}(b). The physical response is largely driven by the self-consistent lattice physics encoded in the Jacobian. As proven in Eq.~\ref{eq:dsigmadU}, $\Sigma(\tau)$ depends monotonously on $U$. $G_0(\tau=\beta/2)$ and $G(\tau=\beta/2)$ also appear to be monotonous in $U$, although the formal proof only holds for $\Sigma$. However, Fig.~\ref{fig:beta20}(c) clearly shows that not all quantities are monotonous in $U$: the Legendre coefficients of the self-energy have a minimum at intermediate $U$.

\begin{figure}
    \includegraphics{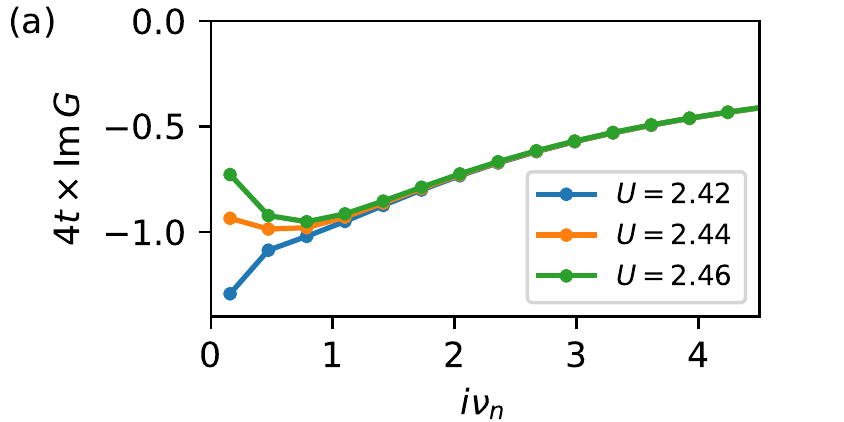}
    \includegraphics{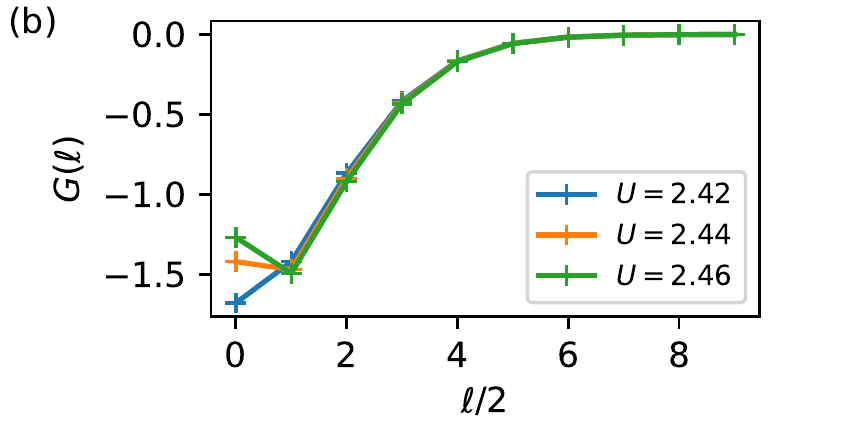}
    \caption{The Green's function in the Matsubara and Legendre representation at $T=1/20$, just above $T_c$ and close to $U_c$. The shape of the curves at low frequency changes rapidly with $U$.}
    \label{fig:criticalpoint}
\end{figure}

Fig.~\ref{fig:criticalpoint}(a) shows the Matsubara Green's function for several value of $U$ close to the critical point. The temperature is $T=1/20>T_c$, as in Fig.~\ref{fig:beta20}. The Green's function has been rescaled by $4t$ here, showing that $G(i\nu_0)\approx G(i\nu_1)\approx 1/(4t)$.
It is directly visible that the slope of $G$ changes as a function of $U$, which is a reflection of the disappearing spectral weight around the Fermi level. 
Fig.~\ref{fig:criticalpoint} shows the Legendre representation of the Green's function. Only even Legendre coefficients are shown, since odd coefficients vanish by particle-hole symmetry. Only the lowest Legendre coefficients differ meaningfully between $U=2.42$ and $U=2.46$. In that sense, the mathematically efficient~\cite{Boehnke11,Dong20} Legendre representation also provides a compact description of the metal-insulator transition.

At $U=2.44$ and $\beta=20$, the symmetrized Jacobian matrix $X^{-1}JX$ is
\begin{gather*}
\begin{pmatrix}
\hphantom{+}0.953& \hphantom{+}0.080& \hphantom{+}0.019& \hphantom{+}0.007& \hphantom{+}0.003& \hphantom{+}0.001 \\
\hphantom{+}0.080& \hphantom{+}0.230& \hphantom{+}0.039& \hphantom{+}0.012& \hphantom{+}0.005& \hphantom{+}0.002 \\
\hphantom{+}0.019& \hphantom{+}0.039& \hphantom{+}0.068& \hphantom{+}0.020& \hphantom{+}0.007& \hphantom{+}0.003 \\
\hphantom{+}0.007& \hphantom{+}0.012& \hphantom{+}0.020& \hphantom{+}0.011& \hphantom{+}0.011& \hphantom{+}0.004 \\
\hphantom{+}0.003& \hphantom{+}0.005& \hphantom{+}0.007& \hphantom{+}0.011& -0.010& \hphantom{+}0.006 \\
\hphantom{+}0.001& \hphantom{+}0.002& \hphantom{+}0.003& \hphantom{+}0.004& \hphantom{+}0.006& -0.018 \\
\end{pmatrix}.
\end{gather*}
The lowest Matsubara frequencies correspond to the top-left of this matrix, and only the first $6\times 6$ matrix elements are shown. The off-diagonal matrix elements, coming from the vertex, are positive and decay quickly with frequency. The diagonal elements start out positive and become negative for larger Matsubara frequencies. The top-left matrix element is close to one, so the leading eigenvector is heavily localized on the lowest Matsubara frequency, but it necessarily has finite weight on the other frequencies as well. 

\section{Conclusion}

Via the susceptibility, a single DMFT calculation provides information about the immediate surroundings of the solution: the iterative stability of solutions, the response to changing parameters and the free energy landscape. The relation between these properties is particularly simple for the Bethe lattice. Here, it was shown that the eigenvalues of the Jacobian have to change sign between the non-interacting and atomic limit, where the Jacobian is dominated by the one-particle and the vertex part, respectively. At the boundaries of the hysteresis region, including the critical end-point, the leading eigenvalue of the Jacobian reaches unity, which leads to a divergent response $d\Sigma/dU$. 

The IPT approximation breaks some of the symmetries of the vertex, and there does not exist a consistent IPT free-energy functional. This shows one of the limitations of using an approximate impurity solver. Still, the vertex-based analysis of the iterative stability and response remains applicable. It is even possible to prove monotonicity of the self-energy, $d\Sigma(\tau)/dU<0$, for any stable IPT solution. 

The IPT vertex has been derived explicitly. At particle-hole symmetry, the vertex has a very simple frequency structure. This explicit expression could form the basis for further analytical considerations of, e.g., the IPT critical point. 
It is noteworthy that the irreducible vertex of the impurity model was not needed in this discussion of the metal-insulator transition on the two-particle level, since all derivations were done with $G_0$ or $\Delta$ and not with $G$. Although zero eigenvalues appear in the Jacobian, the inversion of a local Bethe-Salpeter equation and the associated divergences~\cite{Schafer13,Chalupa18} are avoided in the current formalism.

\section{Outlook}

Efficient, alternative representations of the many-body Green's function have received considerable attention recently~\cite{Boehnke11,Shinaoka17,Li20}. For an overview, see Ref.~\cite{Shinaoka21}. The motivation for these studies is the inefficiency of the Matsubara representation, i.e., the slow, algebraic decay of $G(i\nu_n)$ and $\Sigma(i\nu_n)$. For the finite temperature metal-insulator transition, the relevant physical information distinguishing between a metal and an insulator is largely concentrated at a small number low frequencies, supporting the idea that more compact representations are possible and useful. Indeed, few Legendre coefficients are sufficient to describe the physical changes close to the critical point. Ideally, the representation used for the Green's function should closely match the eigenvectors of the Jacobian, since the magnitude of the eigenvalues of the Jacobian provides a useful criterion for physically relevant and irrelevant degrees of freedom. The Matsubara representation is in this sense efficient (only) in the two trivial limits of the Hubbard model, $U\approx 0$ and $U\gg t$, since the Jacobian is frequency-diagonal in these two cases. 

The DMFT self-consistency conditions were solved using forward iteration here. Other, faster schemes have been used~\cite{Zitko09,Strand11}, which are based on the iterative determination of the Jacobian. The exact DMFT Jacobian, Eq.~\eqref{eq:jacobian}, can be used instead. It requires the vertex, which is generally not cheap to calculate, but a good approximation of the exact vertex, e.g., the IPT vertex or a single-boson exchange formula~\cite{Krien19}, could already help speed up the convergence. Furthermore, only the lowest Matsubara frequencies of the vertex appear relevant for the Jacobian. Similar efficient mixing schemes have also been used in other electronic structure applications~\cite{Pulay80,Kaufmann21,Krien21}.

\begin{acknowledgments}
The author would like to acknowledge useful discussions with Niklas Witt, Friedrich Krien, Emanuel Gull, Koen Reijnders and Hugo Strand.
This work is supported by the Gyllenstierna Krapperup's Foundation. Some of the computations were enabled by resources provided by the Swedish National Infrastructure for
Computing (SNIC) at LUNARC partially funded by the Swedish Research Council through grant agreement no. 2018-05973.
\end{acknowledgments}

\bibliography{references}

\appendix 

\section{Asymptotic expressions}
\label{app:tails}

Here, a brief overview is given of some known properties of many-body Green's functions at large frequencies. More rigorous mathematical treatment can be found elsewhere in the literature~\cite{Hugel16,Rohringer16}.

The Matsubara functions $G(i\nu_n)$, $G_0(i\nu_n)$ and $\Sigma(i\nu_n)$ decay algebraically at large frequency, e.g.,
\begin{align}
    G(i\nu_n) \overset{\nu_n\rightarrow \infty}{\rightarrow} G(M\!=\!0)+\frac{G(M\!=\!1)}{(i\nu_n)}+\frac{G(M\!=\!2)}{(i\nu_n)^{2}}+\ldots, 
\end{align}
where the coefficients $G(M)$ are called the tail coefficients of $G$. At particle-hole symmetry, $\Re G(i\nu_n)=0$, so all even tail coefficients vanish and the lowest-order relevant coefficients are $G(M=1)$ and $G(M=3)$. These coefficients are related to the moments of the (real-energy) spectral function, for example $G(M=1)$ gives the integral of the spectral function and $G(M=3)$ gives its second moment. 

In the imaginary time representation, the first tail coefficient is responsible for the discontinuity at $\tau=0$, 
\begin{align}
    G(M=1)
    &=\lim_{n\rightarrow \infty} i\nu_n G(i\nu_n) \notag \\ 
    &= \lim_{n\rightarrow\infty} i\nu_n \int_0^\beta d\tau G(\tau) e^{i\nu_n\tau} \notag\\
    &= \lim_{n\rightarrow\infty} \int_0^\beta d\tau G(\tau) \frac{d}{d\tau} e^{i\nu_n\tau} \notag\\
    &= G(\tau)e^{i\nu_n\tau}\bigg\vert_{0^+}^{\beta^-} -\lim_{n\rightarrow\infty} \int_0^\beta d\tau \frac{dG(\tau)}{d\tau}   e^{i\nu_n\tau} \notag\\
    &= -1 \cdot G(\tau=\beta^-)-G(\tau=0^+) \notag\\
    &= G(\tau=0^-)-G(\tau=0^+)
\end{align}
Here, $e^{i\nu_n\beta}=-1$ was used and it is argued that the second term in the integration by parts vanishes in the limit of large frequency, since $dG/d\tau$ is smooth on $(0,\beta)$. 

Similarly, the third coefficient is related to the discontinuity in the second derivative of $G(\tau)$ at $\tau=0$, and so on. This can be proven via integration by parts, similar to the proof above. The vanishing of the even coefficients in the particle-hole symmetric system is equivalent to the continuity of the odd derivatives of $G(\tau)$ at $\tau=0$.

From the DMFT-IPT relations, it is possible to derive some of the lowest-order tail coefficients exactly. The self-consistency equation $G_0 = 1/(i\nu_n-t^2 G)$ implies $G_0(M=1)=1$ and $G_0(M=3)=t^2 G(M=1)$ for the self-consistent solution. Dyson's equation $G=G_0/(1-\Sigma G_0)$ implies $G(M=1)=G_0(M=1)$ and $G(M=3)=G_0(M=3)+G_0(M=1)\Sigma(M=1)G_0(M=1)$. Finally, the first moment of $\Sigma$ in IPT is most easily derived in imaginary time, since $G_0(M=1)$ combined with particle-hole symmetry leads to $G_0(\tau=0^-)=1/2$ and $G_0(\tau=0^+)=-1/2$
\begin{align}
    \Sigma^\text{IPT}(M=1) &= \Sigma(\tau=0^-)-\Sigma(\tau=0^+) \notag \\
    &= U^2 \left[G_0(\tau=0^-)^3-G_0(\tau=0^+)^3 \right] \notag \\
    &= \frac{U^2}{4}.
\end{align}
Note that the latter relation also holds for the exact self-energy at particle-hole symmetry, not just for the IPT approximation.
Collecting the results,
\begin{align}
    G_0(i\nu_n) &\overset{\nu_n\rightarrow \infty}{\rightarrow} \frac{1}{i\nu_n} + \frac{t^2}{(i\nu_n)^3}+\ldots \\
    G(i\nu_n) &\overset{\nu_n\rightarrow \infty}{\rightarrow} \frac{1}{i\nu_n} + \frac{t^2+U^2/4}{(i\nu_n)^3}+\ldots \\
    \Sigma(i\nu_n) &\overset{\nu_n\rightarrow \infty}{\rightarrow} \frac{U^2}{4}\frac{1}{i\nu_n} +\ldots ,
\end{align}
shows that the first few tail coefficients can be expressed entirely in terms of the \emph{parameters} $t$ and $U$. This has the important implication that any co-existing solutions in the hysteresis region have the same asymptotic coefficients and therefore differ only ``at small frequencies'', in the sense that $\Sigma^\text{metal}-\Sigma^\text{insulator}$ decays at least as $(i\nu_n)^{-3}$ and $G_0^\text{metal}-G_0^\text{insulator}$ decays at least as $(i\nu_n)^{-5}$.

On the other hand, the third moment of the IPT self-energy,
\begin{align}
    \Sigma^\text{IPT}(M=3) &= \frac{d^2\Sigma(\tau)}{d\tau^2}\bigg\vert_{0^+}^{0^-} \\
    &= \left(3U^2 G_0^2 \frac{d^2 G_0(\tau)}{d\tau^2} + 6U^2 G_0 \left(\frac{dG_0}{d\tau}\right)^2 \right)\bigg\vert_{0^+}^{0^-} \notag \\
    &=\frac{3U^2}{4} G_0(M=3) + 6 U^2  \left(\frac{dG_0}{d\tau}\big\vert_{\tau=0}\right)^2. \notag
\end{align}
This follows directly from the IPT expression for the self-energy, Eq.~\eqref{eq:ipt}, at particle-hole symmetry.
Here, the actual value of the first derivative of $G(\tau)$ at $\tau=0$ enters, not just the discontinuity. Thus, the right-hand side of this expression cannot be expressed entirely in terms of the parameters and can be different for co-existing solutions of the self-consistency relation. All results in this Appendix also hold for the exact DMFT solution on the particle-hole symmetric Bethe lattice.

\section{Curie-Weiss mean-field theory}
\label{app:curieweiss}

Curie-Weiss mean-field theory for the spin-1/2 Ising model is perhaps the most familiar example of a mean-field theory. Reducing the notation to a minimum, it is defined by the set of equations
\begin{align}
M(h) &= \tanh(\beta h),\\
h&=B+\alpha M.
\end{align}
Here, $M(h)$ is the magnetization of the impurity model as a function of the self-consistent field $h$ and $h=B+\alpha M$ is the self-consistency relation. The physical variables are the inverse temperature $\beta$ and the external field strength $B$, $\alpha$ is a fixed constant. Clearly $h=0$ and $M=0$ is always a solution when $B=0$.

The stability of this solution is given by the ``Jacobian'', $J=dM^\text{new}/dM^\text{old}=dM/dh \cdot  \partial h/\partial M =\alpha \, dM/dh =\alpha \beta/\cosh^2(\beta h)$. Filling in $h=0$ gives $J=1$ if and only if $\alpha \beta=1$. In other words, $\beta_c=1/\alpha$. 

The physical susceptibility $dM/dB$ can be written as
\begin{align}
    \frac{dM}{dB} &= \frac{dM}{dh} \frac{dh}{dB} = \frac{dM}{dh} \left(1+\alpha\frac{dM}{dB}\right) \\
    \frac{dM}{dB} &= \frac{dM/dh}{1-\alpha \, dM/dh }.
\end{align}
This diverges at the critical point $J=1$, since the denominator is equal to zero there. This illustrates the relation between the Jacobian and the response function in Curie-Weiss mean-field theory.

\section{Representations of the IPT vertex}
\label{app:vertexreps}

In the main text, the reducible vertex of IPT is given in imaginary time and in imaginary frequency. To see how these two expression are related, it is useful to derive the imaginary time result in a slightly more general set-up. In the main text, time translation symmetry is assumed throughout the derivation and $\Sigma$ and $G$ are written as a function of a single imaginary time argument. This implicitly sets the bosonic frequency $\omega$ of the vertex to zero.

Instead, it is also possible to start from the two-time expression
\begin{align}
    \Sigma^\text{IPT}(\tau_1,\tau_2) &= -U^2 G_0(\tau_1,\tau_2)G_0(\tau_2,\tau_1)G_0(\tau_1,\tau_2).
\end{align}
For the vertex, this gives a four-time object
\begin{align}
    \frac{\partial \Sigma(\tau_1,\tau_2)}{\partial G_0(\tau_3,\tau_4)} =& -2 U^2 \delta_{\tau_1,\tau_3}\delta_{\tau_2,\tau_4}G_0(\tau_1,\tau_2)G_0(\tau_2,\tau_1)\notag \\
    &-U\delta_{\tau_2,\tau_3}\delta_{\tau_1,\tau_4}G_0(\tau_1,\tau_2)G_0(\tau_1,\tau_2).
\end{align}
Now, when considering variations around the time translation symmetric solution, time translation symmetry can be used on the right-hand side, i.e., $G_0(\tau_1,\tau_2)=G_0(\tau_2-\tau_1)$. Particle-hole symmetry gives $G_0(\tau)=-G_0(-\tau)$. Together, this leads to
\begin{align}
    \frac{\partial \Sigma(\tau_1,\tau_2)}{\partial G_0(\tau_3,\tau_4)} =& 2 U^2 \delta_{\tau_1,\tau_3}\delta_{\tau_2,\tau_4}G^2_0(\tau_2-\tau_1)\notag \\
    &-U\delta_{\tau_2,\tau_3}\delta_{\tau_1,\tau_4}G^2_0(\tau_2-\tau_1).
\end{align}
Clearly, every part of this formula only depends on time differences. The appearance of two $\delta$ functions means that both terms depend on a single linear combination of frequencies only. 

For the vertex in the frequency representation, a more compact expression is obtained by defining a ``bare susceptibility'' $\chi$ as
\begin{align}
    \chi(i\omega) &= -\sum_{\nu_1} G_0(i\nu_1) G_0(i\nu_1-i\omega), \label{eq:susc}\\
    \delta \Sigma(\nu)/\delta G_0(\abs{\nu'}) &= 3 \frac{U^2}{\beta^2} \left[\chi(i\nu-i\nu')-\chi(i\nu+i\nu')\right].\notag
\end{align}
Differences and sums of fermionic Matsubara frequencies appear here, which results in bosonic Matsubara frequencies, e.g., $\nu_i-\nu_j=(2i-2j)\pi T=\omega_{i-j}$ and $\nu_i+\nu_j=(2i+2j+2)\pi T=\omega_{i+j+1}$. An important detail is the $+1$ in the last expression. Note that $\chi$ is real and positive, since $G_0$ is purely imaginary.

It is easy to verify that $\chi(-i\omega)=\chi(i\omega)$, by relabelling $\nu_1\rightarrow \nu_2=\nu_1-\omega$, so the expression for $\partial \Sigma(i\nu)/\partial G_0(\abs{\nu'})$ is properly antisymmetric in $\nu$. This confirms that $\Sigma$ remains particle-hole symmetric as long as $G_0$ is particle-hole symmetric. 

The frequency structure of the vertex is given entirely by $\chi(i\nu-i\nu')-\chi(i\nu+i\nu')$, with $3U^2/\beta^2$ appearing as a simple prefactor. At particle-hole symmetry, it is sufficient to consider the matrix $\delta \Sigma(i\nu)/\delta G_0(\abs{\nu'})$ for $\nu>0$ and $\nu'>0$. This matrix is real and symmetric. The first contribution, $\chi(i\nu-i\nu')$, depends only on the distance to the diagonal of the matrix (Toeplitz matrix), whereas the second contribution, $\chi(i\nu+i\nu')$ depends only on the Manhattan distance to the top-left corner of the matrix (Hankel matrix), see Fig.~\ref{fig:matrix}. Note that that Hartree term in the self-energy would provide a term proportional to $U^1$ in the vertex, but this term would be symmetric in the fermionic frequency and at particle-hole symmetry only the frequency-antisymmetric part plays a (non-trivial) role. 

The IPT vertex simplifies substantially in the atomic limit~\cite{Thunstrom18}, where $G_0=1/i\nu_n$, since
\begin{align}
    \chi(\omega_0) &= \frac{1}{\pi^2 T^2} \sum_n \frac{1}{2n+1}\frac{1}{2n+1} = \frac{1}{4 T^2},
\end{align}
and for finite frequency $a\neq 0$, manipulations similar to the usual evaluation of the Lindhard bubble give
\begin{align}
    G_0(i\nu_n)G_0(i\nu_{n+a}) &= \frac{ G_0(i\nu_n) - G_0(i\nu_{n+a})}{ G_0^{-1}(i\nu_{n+a})- G_0^{-1}(i\nu_n) } \notag \\
    &=\frac{ G_0(i\nu_n) - G_0(i\nu_{n+a})}{ 2ai }, \notag  \\
    \chi(\omega_a) &= \sum_n G_0(i\nu_n)G_0(i\nu_{n+a}) \notag  \\
    &= \frac{1}{2ai} \left(\sum_n G_0(i\nu_n) - \sum_n G_0(i\nu_{n+a}) \right) \notag \\
    &= 0.
\end{align}
This reflects the fact that the atomic limit does not have any charge dynamics. Looking at Eq.~\eqref{eq:dSigmadG0:antisymmetrized} or at Fig.~\ref{fig:matrix}, the vertex is proportional to the identity matrix. Note that, although $\Sigma^\text{IPT}=\Sigma^\text{Exact}$ at $t=0$, this does not imply that the corresponding vertices are also equal, so the proportionality of the vertex with the identity matrix is a result that is derived within the IPT approximation. For a detailed investigation of the atomic limit, see Ref.~\cite{Thunstrom18}. 

Going back to the general situation, it is interesting to point out that $\chi(\omega_0)=\norm{G_0}^2$ according to Eq.~\eqref{eq:norm}. For finite $\omega_n$,
\begin{align}
    \chi(\omega_n) &= \int_0^\beta d\tau \cos(\omega_n \tau) G_0^2(\tau), 
\end{align}
where symmetry was used to obtain the $\cos$ instead of a complex exponential. Since $G_0^2(\tau)$ is positive-definite and $\cos(\omega_n \tau)\leq 1$, this implies that $0\leq \chi(\omega_n) \leq \chi(\omega_0)$. Numerically it turns out that $\chi(\omega_n)$ is a decreasing function of $n\geq 0$ (for causal $G_0$, the Lehmann representation provides a way to prove this). This implies that all matrix elements of the IPT vertex are positive, since they are of the form $\chi(\omega_a)-\chi(\omega_b)$ with $b>a\geq0$. As seen above, in the atomic limit $\chi(\omega_n)=0$ for $n>0$, so the matrix elements are positive but not strictly positive. A further remark is that the mapping between $G_0$ and the vertex is one-to-one, given fixed values of $U$ and $\beta$.

The structure of Fig.~\ref{fig:matrix} shows that the rows of the vertex are telescoping, leading to $R_j=3U^2/\beta^2 \left[\chi(i\omega_0)+2\sum_{m=1}^j \chi(i\omega_m)\right]$ for the sum over the $j$-th row. This row sum enters the Gershgorin circle theorem~\cite{Gershgorin} for the eigenvalues of a matrix. It is plausible that further analytical results for the eigenvalues of the IPT vertex can be derived given suitable assumptions on $G_0$. 

\begin{figure}
    \begin{tikzpicture}[square/.style={regular polygon,regular polygon sides=4}]
    
    \node at (1,1) {$\chi(i\nu-i\nu')$}; 
    
    \fill[matrix0] (-0.5,0.5) rectangle ++(1,-1);
    \fill[matrix0] (1.5,-1.5) rectangle ++(1,-1);
    \fill[matrix0] (0.5,-0.5) rectangle ++(1,-1);
    
    \fill[matrix1] (0.5,0.5) rectangle ++(1,-1);
    \fill[matrix1] (-0.5,-0.5) rectangle ++(1,-1);
    \fill[matrix1] (1.5,-0.5) rectangle ++(1,-1);
    \fill[matrix1] (0.5,-1.5) rectangle ++(1,-1);
    
    \fill[matrix2] (1.5,0.5) rectangle ++(1,-1);
    \fill[matrix2] (-0.5,-1.5) rectangle ++(1,-1);

     \node[] at (0,0) {$\chi(\omega_0)$} ;
     \node[] at (1,0) {$\chi(\omega_1)$} ;
     \node[] at (2,0) {$\chi(\omega_2)$} ;
     \node[] at (0,-1) {$\chi(\omega_1)$} ;
     \node[] at (1,-1) {$\chi(\omega_0)$} ;
     \node[] at (2,-1) {$\chi(\omega_1)$} ;
     \node[] at (0,-2) {$\chi(\omega_2)$} ;
     \node[] at (1,-2) {$\chi(\omega_1)$} ;
     \node[] at (2,-2) {$\chi(\omega_0)$} ;

\draw [decorate,decoration={brace,amplitude=10pt},yshift=0pt] (-0.5,1.5) -- (6.5,1.5) ;
\node at (3,2.5) {$\dfrac{\beta^2}{3U^2}\dfrac{\partial \Sigma}{\partial G_0}$}  ;

\node at (3,-1) {$-$}  ;
\begin{scope}[shift={(4,0)}]
 
    \node at (1,1) {$\chi(i\nu+i\nu')$}; 
    
    \fill[matrix1] (-0.5,0.5) rectangle ++(1,-1);
    
    \fill[matrix2] (0.5,0.5) rectangle ++(1,-1);
    \fill[matrix2] (-0.5,-0.5) rectangle ++(1,-1);
    
    \fill[matrix3] (1.5,0.5) rectangle ++(1,-1);
    \fill[matrix3] (0.5,-0.5) rectangle ++(1,-1);
    \fill[matrix3] (-0.5,-1.5) rectangle ++(1,-1);

    \fill[matrix4] (1.5,-0.5) rectangle ++(1,-1);
    \fill[matrix4] (0.5,-1.5) rectangle ++(1,-1);

    \fill[matrix5] (1.5,-1.5) rectangle ++(1,-1);

     \node[] at (0,0) {$\chi(\omega_1)$} ;
     \node[] at (1,0) {$\chi(\omega_2)$} ;
     \node[] at (2,0) {$\chi(\omega_3)$} ;
     \node[] at (0,-1) {$\chi(\omega_2)$} ;
     \node[] at (1,-1) {$\chi(\omega_3)$} ;
     \node[] at (2,-1) {$\chi(\omega_4)$} ;
     \node[] at (0,-2) {$\chi(\omega_3)$} ;
     \node[] at (1,-2) {$\chi(\omega_4)$} ;
     \node[] at (2,-2) {$\chi(\omega_5)$} ;
 \end{scope}
    \end{tikzpicture}    
    \caption{The IPT vertex is given by the difference between a Toeplitz matrix (left) and a Hankel matrix (right), and a prefactor $3U^2/\beta^2$.}
    \label{fig:matrix}
\end{figure}

The IPT vertex is very simple, since it consists of a single diagram at half-filling. It is not necessary to resort to channel decomposition techniques~\cite{Krien19,Wentzell20}, since this decomposition in terms of a ``susceptibility'' occurs naturally. Unlike the usual channel decompositions, the ``susceptibility'' that appears here is based on $G_0$ and not on $G$. 

\section{Zero temperature}
\label{app:T0}

Here, the limit $T\rightarrow 0$ of IPT for the Hubbard model is considered, in combination with the two limits of $U\gg t$ and $U\approx 0$. 

\emph{Atomic limit:}
for $t=0$, $G_0(i\nu)=1/i\nu$, $\Sigma(i\nu)=U^2/(4i\nu)$ and $\chi(i\omega_m)=\frac{1}{4T^2}\delta_{m,0}$.
This gives $G^{-1}(i\nu_0)=G_0^{-1}(i\nu_0)-\Sigma(i\nu_0) \approx -\Sigma(i\nu_0)$. 
The leading eigenvalue of the Jacobian is determined by the diagonal matrix element of the vertex contribution, $\lambda\approx J^\Sigma(\nu_0,\nu_0)$,
\begin{align}
    \lambda &\approx \frac{3U^2 t^2}{\beta^2} G_0^2(i\nu_0) G^2(i\nu_0) \chi(\omega_0) \notag \\
    &\approx \frac{3U^2 t^2}{\beta^2} \frac{1}{\pi^2 T^2} \frac{16\pi^2 T^2}{U^4} \frac{1}{4 T^2} \notag \\
    &= \frac{12 t^2}{U^2} \text{ for large $U$}.
\end{align}

\emph{Non-interacting limit:}
The limit $T\rightarrow 0$ implies that the discrete Matsubara frequencies approach the origin of the complex plane, so $G_0(i\nu)$ is directly related to $G_0(E=0)$ in this limit. At small $U$, $G_0$ stays close to the non-interacting value, in the sense that $\Im G_0(i\nu)\rightarrow -\frac{1}{t}$ for $\nu \approx 0^+$.  For the IPT vertex, the difference of two susceptibilities is needed. The simplest case is 
\begin{align}
    \chi(i\omega_0)-\chi(i\omega_{1}) &= \sum_m G(i\nu_m) \left[G(i\nu_m)-G(i\nu_{m-1})\right] \notag \\
    &\approx G(i\nu_0)\left[G(i\nu_{0})- G(i\nu_{-1}) \right]\notag \\&\phantom{=}+\sum_{m\neq 0} G(i\nu_m) \, 2\pi T \, \frac{dG(i\nu)}{d(i\nu)} \notag \\
    &= 2 G(i\nu_0)^2 + O(T) \notag \\
    &= 2t^{-2} \text{ for $T\rightarrow 0$.}
\end{align}
Here, it was used that at low temperature, the Matsubara grid becomes dense and $G(i\nu_n)$ is a smooth function except between $\nu_{-1}$ and $\nu_{0}$, where it jumps from $+1/t$ to $-1/t$. A similar derivation shows that $\chi(i\omega_m)-\chi(i\omega_{m'})=(m'-m)\cdot 2t^{-2}$ for $m'>m\geq 0$. The structure of Fig.~\ref{fig:matrix} means that the first row and column of the vertex are constant and equal to $3\frac{U^2}{\beta^2} 2t^{-2}$, the second row and column are constant and equal to $3\frac{U^2}{\beta^2}\cdot 3\cdot 2t^{-2}$ except for their first elements, and so on. The factor in front of the vertex $\beta^{-2}=T^2$ makes the low-frequency components of the vertex disappear in the limit $T\rightarrow 0$. Since $G_0(i\nu_0)$ is approximately constant in the limit of low temperature, the vertex contribution $J^\Sigma$ in Eq.~\eqref{eq:jacobian} indeed vanishes compared to $J^0$, and the leading eigenvalue is still given by $t^2 G^2(i\nu_0)$, with the eigenvector concentrated on the lowest Matsubara frequency. So at small $U$, the leading eigenvalue is $\lambda_J \approx -1$. 

\end{document}